
\magnification=\magstep1

\vsize=8.7truein
\hsize=6.1truein
\vskip 0.25in
\baselineskip=14pt plus 0.2pt minus 0.3pt
\parskip=12pt
\parindent=0pt
\hoffset=0.3in
\footline={\ifnum\pageno<2\hfil\else\hss\tenrm\folio\hss\fi}
\font\bigrm=cmr10 scaled\magstep3
\def\H{{\cal H}}
\def\Hi{{\cal H}_i}
\def\frac#1#2{\textstyle{#1\over #2}\displaystyle}
\def\B3{{^3\!B}}
\def\Hthree{{^3\!H}}

\def\S{{\cal S}}
\def\sss{\scriptscriptstyle}
{\bigrm Microcanonical Action and the}
\vskip 0pt
{\bigrm Entropy of a Rotating Black Hole}
\vskip 5pt
{\bigrm J. David Brown}\hfil\break
Departments of Physics and Mathematics\hfil\break
North Carolina State University, Raleigh, NC 27695--8202
\vskip 5pt
{\bigrm James W. York, Jr.}\hfil\break
Department of Physics and Astronomy\hfil\break
University of North Carolina, Chapel Hill, NC 27599--3255

The authors have recently proposed a ``microcanonical functional
integral" representation of the density of quantum states of
the gravitational field. The phase of this real--time functional
integral is determined by a ``microcanonical" or Jacobi action,
the extrema of which are classical solutions at fixed total
energy, not at fixed total time interval as in Hamilton's
action. This approach is fully general but is especially well
suited to gravitating systems because for them the total energy
can be fixed simply as a boundary condition on the gravitational
field. In this paper we describe how to obtain Jacobi's action
for general relativity. We evaluate it for a certain complex
metric associated with a rotating black hole and discuss the
relation of the result to the density of states and to the entropy
of the black hole.

{\bf 1 DEDICATION}\hfil\break
We dedicate this paper to Yvonne Choquet--Bruhat in honor of her
retirement following a brilliant career. JWY would like to thank
her for friendship, support, and their happy collaboration
in ``analysis, manifolds, and physics" [1].

{\bf 2 INTRODUCTION}\hfil\break
We are concerned here with the description of a stationary and
axisymmetric rotating black hole in a closed system in thermodynamic
equilibrium with its environment. The equilibrating radiation in the
closed system surrounding the hole is the ``environment". This
``radiation fluid" rotates at a constant angular velocity [2][3].
Unlike the more familiar non--relativistic case, here one must
take into account explicitly the spatial finiteness of the
system in order to avoid super--luminal velocities resulting from
the rotation. Even though we shall ignore the explicit effects of
the equilibrating radiation in this work, its presence--in--principle
must be kept in mind in order to give the problem a physically
and mathematically reasonable formulation.

The key conserved quantities in closed systems like ours are the
total energy and the total angular momentum. Their ``Massieu"
conjugates [4] are, respectively, inverse temperature $\beta$, and
$\beta$ multiplied by an appropriate angular velocity. An important
question is which among these quantities to specify in advance,
that is, which ``ensemble" picture to employ. We choose here to
fix energy and angular momentum as in the microcanonical ensemble.
This choice leads us to employ the general relativistic version
of Jacobi's action.

Jacobi's form of the action principle [5][6] involves variations
at fixed energy, rather than the variations at fixed time used
in Hamilton's principle. The fixed time interval in Hamilton's
action becomes fixed inverse temperature $\beta$ in the ``periodic
imaginary time" formulation, thus transforming Hamilton's action
into the appropriate (imaginary) phase for a periodic path in
computing the canonical partition function from a
``Euclidean" Feynman functional integral [7]. (We are here and
in the next paragraph speaking only of energy and inverse
temperature in order to simplify the discussion. Similar
remarks hold for angular momentum and its conjugate.) In contrast,
fixed energy is suitable for the microcanonical ensemble and,
correspondingly, Jacobi's action is the phase in an expression
for the density of states as a real--time ``microcanonical
functional integral" (MCFI) [8].

Let us characterize briefly the canonical and microcanonical
pictures. Neither picture can hold with perfect precision even in
principle when gravity is taken into account because the ``infinite"
heat bath of the canonical picture and the ``perfectly adiabatic"
walls of the microcanonical picture are both at variance with the
known physics of the gravitational field. However, each picture
still provides a useful framework for discussion. Furthermore, the
long--range, unscreened nature of the gravitational interaction
means that simple notions of ordinary statistical thermodynamics
like ``extensive", ``intensive", {\it etc.\/}, no longer apply
in general. With gravity, statistical mechanics is inherently
{\it global\/}.

In the canonical picture, with a fixed temperature shared by all
constituents of a system, there are no constraints on the energy.
This feature simplifies combinatorial (counting) problems and leads
to factorization of the partition function for weakly coupled
constituents. For gravitating systems in equilibrium, the
temperature is not spatially uniform because of red--shift and
blue--shift effects. In such cases, the relevant temperature
is that determined at the spatial two--boundary $B$ of the system
[9]. It can be specified by a boundary condition on the metric
[10][11]. It is then used in conjunction with Hamilton's principle,
which is the form of the gravity action in which the metric is
fixed on the history of the spatial two--boundary [12]. (The metric
determines the lapse of proper time along the history of the
boundary.) The case of a rotating, charged, stationary, axisymmetric
black hole has been treated with (grand) canonical boundary
conditions in [13]. See also [10] and [11].

With its constraint on the energy, the microcanonical picture leads
to stability properties more robust than in the canonical case.
However, the energy constraint can complicate combinatorial problems
because the constituents of the system must all share from a common
fixed pool of energy. For field theories, with a continuous infinity
of degrees of freedom, the energy constraint restricts  the entire
phase space of the system {\it unless gravity is taken into
account\/}. For gravitating systems, as a  consequence of the
equivalence principle, the total energy in a given finite region,
including that of matter fields, can be given as an integral of
certain geometrically well--defined derivatives of the metric over
a finite two--surface bounding the system. In other words, one
{\it can\/} find and employ a suitable expression for ``quasi--local"
gravitational energy that does not require an appeal to asymptotic
regions. This quasi--local energy has the value, per unit proper
time, of the Hamiltonian of the spatially bounded region, as
discussed in detail in [12]. Therefore, if we specify as a boundary
condition the energy per unit two--surface area, we have constrained
the total energy simply by a boundary condition [8][12]. Thus,
through the mediation of the gravitational field, the canonical and
microcanonical cases are placed on similar footing and differ only
in which of the conjugate variables [14], inverse temperature or
energy, is specified on the boundary. The corresponding functional
integrals, for the partition function or density of states, differ
in which action gives the correct phase, Hamilton's or Jacobi's.
We regard the MCFI for the density of states as the more fundamental
[8].

We shall now outline a recent application of the above reasoning
to the case of an axisymmetric stationary black hole. The MCFI,
in a steepest descents approximation, shows that the density of
states is the exponential of one--fourth of the area of the event
horizon, thus confirming in this approximation the
Bekenstein--Hawking expression for the entropy of a black hole [8].
Full details of the following are given in [8] and [12]. (An
application of the MCFI to a simple harmonic oscillator--without
gravity--has been given recently by the authors [15]. No
approximations were required in this case and the exact energy
spectrum was obtained.)

{\bf 3 JACOBI'S ACTION FOR GRAVITY}\hfil\break
We here analyze the action for pure general relativity. The method,
however, can be just as well applied when matter is included and/or
for higher--derivative theories of gravity.

Consider a region of spacetime $M=\Sigma\times I$ where $\Sigma$ are
spatial slices and $I$ is an interval of the real line. The
two--boundary of space $\Sigma$ is denoted by $B$, whose history is
$\B3 = B\times I$. The orthogonal intersection of any $\Sigma$ with
$\B3$ is the two--boundary $B$ at some time. Thus a generic $B$
can be regarded as being embedded in $\Sigma$ with a spacelike unit
normal $n^\mu$ tangent to $\Sigma$ at $B$, and also as being embedded
in $\B3$ with a timelike unit normal $u^\mu$ such that
$u_\mu n^\mu =0$. The subspaces of $M$ that correspond to the
endpoints of $I$ are spacelike hypersurfaces $t=t'$ and $t=t''$.
The notation $\int^{t''}_{t'} d^3x$  denotes an integral over $t''$
minus an integral over $t'$. The spacetime metric is $g_{\mu\nu}$ and
its scalar curvature is $\Re$. Then ``Hamilton's" action for general
relativity, in which the metric is fixed on the boundary, is given
by a variant of the Hilbert action, namely [12][16]
$$ S[g] = {1\over 2\kappa} \int_M d^4x\sqrt{-g} (\Re - 2\Lambda)
  + {1\over\kappa}\int^{t''}_{t'} d^3x\sqrt{h} K - {1\over\kappa}
   \int_\B3   d^3x\sqrt{-\gamma}\Theta  - S^0 \ ,\eqno(1)$$
where the metric and extrinsic curvature of the slices $\Sigma$ are
$h_{ij}$ and $K_{ij}$, while those for $\B3$ are $\gamma_{ij}$ and
$\Theta_{ij}$. (Latin letters $i$, $j$, $\ldots$ are used as tensor
indices for both $\Sigma$ and $\B3$. No cause for confusion arises
from this convention.) In (1), $\kappa=8\pi G$ and we set Newton's
constant $G=1$ henceforth. The term $S^0$ in (1) is a functional of
the metric $\gamma_{ij}$ of $\B3$. The purpose of this term is to
determine the ``zero" of energy and momentum. It will not turn out
to affect the Jacobi action for general relativity. ($S^0$ is
analyzed in [12] but was not included in [16].)

The canonical form of ``Hamilton's" action (1) is [12]
$$S = \int_M d^4x \bigl[ P^{ij} {\dot h}_{ij} - N\H - V^i\Hi \bigr]
   -  \int_\B3 d^3x
  \sqrt{\sigma} \bigl[ N\varepsilon - V^i j_i \bigr] \ ,\eqno(2)$$
where $P^{ij}$ is the Arnowitt--Deser--Misner momentum conjugate to
$h_{ij}$, and $\H$ and $\Hi$ are the usual Hamiltonian and momentum
constraints. The lapse function is denoted by $N$ and the shift
vector by $V^i$. In the surface term of (2), $\sigma$ denotes the
determinant of the two--metric $\sigma_{ij}$ induced on $B$ as a
surface embedded in $\Sigma$; likewise, $n_i$ denotes $B$'s unit
normal in $\Sigma$ and $k_{ij}$ denotes the corresponding extrinsic
curvature. The energy surface density $\varepsilon$ and momentum
surface density $j_i$ are given by [12]
$$\eqalign{\varepsilon &= {1\over\kappa}k + {1\over\sqrt{\sigma}}
                            {\delta S^0\over\delta N} \ ,\cr
             j_i &= -{2\over\sqrt{h}}\sigma_{ij}n_kP^{jk} -
                    {1\over\sqrt{\sigma}}
                    {\delta S^0\over\delta V^i}\ .\cr}\eqno(3)$$
The total quasi--local energy is the integral of $\varepsilon$ over
$B$. In obtaining (2), we have assumed that $S^0$, if present, is a
linear functional of $N$ and $V^i$ on $\B3$. In the total Hamiltonian
extracted directly from (2), the shift vector must satisfy
$n_i V^i|_B=0$. Each of these points is discussed in [12].

In the action (2), one varies $h_{ij}$, $P^{ij}$, $N$, and $V^i$ to
obtain
$$\eqalignno{ \delta S &= \,({\hbox{terms giving the equations of
     motion}})   +\int^{t''}_{t'} d^3x \, P^{ij}\delta h_{ij} \cr
  &\quad - \int_\B3 d^3x\sqrt{\sigma} \Bigl[ \varepsilon\,\delta N
        - j_a  \delta V^a - (N/2)s^{ab}\delta\sigma_{ab} \Bigr]
            \ ,&(4)\cr}$$
where indices $a$, $b$, $\ldots$ are used to denote tensors on $B$,
{\it i.e.\/}, $\Sigma$--tensors that are orthogonal to $n^i$. The
surface stress tensor $s^{ab}$ on $B$ is given in [12] and does
not concern us here. We see from (4) that suitable boundary
conditions for $S$ are obtained by fixing the metric induced on
the boundary elements $t'$, $t''$, and $\B3$ of $M$. In particular,
the lapse function $N$ is fixed on $\B3$, where it determines
proper time elements $N dt$ on $\B3$ along the unit normal $u^\mu$
associated with the foliation of $\B3$ by $B$; hence, we refer to
$S$ as ``Hamilton's action" in canonical form.

What we define as the microcanonical action $S_m$ is, in essence,
Jacobi's action for general relativity. It is
obtained from $S$ by a canonical transformation that changes the
appropriate boundary conditions on $\B3$ from fixed metric components
$N$, $V^a$, and $\sigma_{ab}$ to fixed energy surface density
$\varepsilon$, momentum surface density $j_a$, and boundary metric
$\sigma_{ab}$. Thus, define [8]
$$\eqalign{S_m &= S +\int_\B3 d^3x\sqrt{\sigma} \bigl[ N\varepsilon
    -V^aj_a \bigr] \cr  &= \int_M d^4x \bigl[ P^{ij} {\dot h}_{ij}
     - N\H - V^i\Hi \bigr]   \ .\cr}\eqno(5)$$
{}From (4), it follows that the variation of $S_m$ is
$$\eqalignno{ \delta S_m &= \,({\hbox{terms giving the equations of
    motion}})   +\int^{t''}_{t'} d^3x \, P^{ij}\delta h_{ij} \cr
   &\quad + \int_\B3 d^3x \Bigl[ N\,\delta(\sqrt{\sigma}\varepsilon)
    - V^a  \delta(\sqrt{\sigma} j_a) + (N\sqrt{\sigma}/2)s^{ab}
                 \delta\sigma_{ab} \Bigr]   \ .&(6)\cr}$$
This result shows that solutions of the equations of motion extremize
$S_m$ with respect to variations in which $\varepsilon$, $j_a$, and
$\sigma_{ab}$ are held fixed on the boundary $B$. Observe that the
unspecified subtraction term $S^0$ does not appear in $S_m$.
Nevertheless, the variation (6) of $S_m$ involves $\varepsilon$, $j_a$,
and $s^{ab}$, which do depend on $S^0$ through their definitions.
However, all dependences on $S^0$ in the boundary variation terms of
$\delta S_m$ actually cancel because of the requirement that $S^0$
be a linear functional of the lapse and shift [8][12]. Thus, neither
$S_m$ nor its variation depends on $S^0$. In other words, as long as
energy and momentum are to be fixed, as in $S_m$, a quantity like
$S^0$ whose only role is to determine their ``zero points" is
irrelevant.

{\bf 4 MICROCANONICAL FUNCTIONAL INTEGRAL}\hfil\break
In [8] and [15] we showed that for nonrelativistic mechanics the density
of states is given by a sum over periodic, real time histories, where each
history contributes a phase determined by Jacobi's action. In the case of
nonrelativistic mechanics, the energy that is fixed in Jacobi's action
is just the value of the Hamiltonian
that generates unit time translations. For a self--gravitating system, the
Hamiltonian has a ``many--fingered"  character:  space can be pushed into
the future in a variety of ways, governed by different choices of lapse
function $N$ and shift vector $V^i$. The value of
the Hamiltonian incorporated into (2) depends on this choice. More precisely,
the value of the Hamiltonian is determined by the choice of lapse and shift
on the boundary $B$, since the lapse and shift on the domain of $\Sigma$
{\it interior\/} to $B$ are Lagrange multipliers for the (vanishing)
Hamiltonian and momentum constraints. Accordingly, the energy
surface--density $\varepsilon$ and momentum surface--density $j_a$ for a
self--gravitating system play a role that is analogous to energy for a
nonrelativistic mechanical system.

The above considerations lead us to propose that the density of states for a
spatially finite, self--gravitating system is a functional of the energy
surface--density $\varepsilon$ and  momentum surface--density $j_a$. In
addition to these energy--like quantities, the density of states is also a
functional of the metric $\sigma_{ab}$ on the boundary $B$, which specifies
the size and shape of the system. In the absence of matter fields, these make
up the complete set of  variables and $\nu[\varepsilon,j_a,\sigma_{ab}]$ is
interpreted as the density of quantum states of the gravitational field with
energy density, momentum density, and boundary metric having the values
$\varepsilon$, $j_a$, and $\sigma_{ab}$. The action to be used in the
functional integral representation of $\nu$ is $S_m$, which describes the
gravitational field with fixed $\varepsilon$, $j_a$, and $\sigma_{ab}$. Note
that $\varepsilon$, $j_a$, and $\sigma_{ab}$ replace the traditional
thermodynamical extensive variables. Our variables are all constructed
from the dynamical phase space variables ($h_{ij}$, $P^{ij}$) for the system,
where the phase space structure is defined using the foliation of $M$ into
spacelike hypersurfaces. (We expect this to be a defining feature of
extensive variables for general systems of gravitational and matter fields.)

We propose [8] that the density of states of the gravitational field is
defined formally by
$$\nu[\varepsilon,j,\sigma] = \sum_M \int {\cal D}H \exp(iS_m) \ .\eqno(7)$$
(Planck's constant has been set to unity.) The sum over $M$ refers to a sum
over manifolds of different topologies. The three--boundary for each $M$ is
required to have topology $\partial M = B\times S^1$. If $B$ has two--sphere
topology, then the sum over topologies includes $M=({\hbox{ball}})\times S^1$,
with $\partial M = \partial({\hbox{ball}})\times S^1 = S^2\times S^1$. Another
example is $M=({\hbox{disk}})\times S^2$, with $\partial M =
\partial({\hbox{disk}})\times S^2 = S^1\times S^2$. The action $S_m$ that
appears in Eq.~(7) is the microcanonical action (5) of the previous
section applied to the manifolds $M$ with a single boundary component
$\partial M = \B3 = B\times S^1$.
The functional integral (7) for $\nu$ is a sum over Lorentzian metrics
$g_{\mu\nu}$. Note that the microcanonical action may require the addition
of terms that depend on the topology of $M$, such as the Euler number.

In the boundary conditions on $\partial M = B\times S^1$, the two--metric
$\sigma_{ab}$ that is fixed on the hypersurfaces $B$ is typically real and
spacelike. Likewise, the energy density $\varepsilon$ is real, which requires
the unit normal to $\partial M$ to be spacelike. Therefore, the Lorentzian
metrics on $M$ must induce a Lorentzian metric on $\partial M$, where the
timelike direction coincides with the periodically identified $S^1$. Note,
however, that there are no nondegenerate Lorentzian metrics on a manifold with
topology $M=({\hbox{disk}})\times S^2$ that also induce such a Lorentzian
metric on $\partial M$. This implies that the formal functional integral (7)
for the density of states must include degenerate metrics. (For a discussion
of the role of degenerate metrics in classical and quantum gravity, see [17].)

Now consider the evaluation of the functional integral (7) for fixed
boundary data $\varepsilon$, $j_a$, $\sigma_{ab}$ that correspond to a
stationary, axisymmetric black hole. That is, start with a real Lorentzian,
stationary, axisymmetric, black hole solution of the Einstein equations, and
let $T = {\rm constant}$ be stationary time slices that contain the closed
orbits of the axial Killing vector field. Next, choose a topologically
spherical two--surface $B$ that contains the orbits of the axial Killing
vector field, and is contained in a $T= {\rm constant}$ hypersurface.
{}From this surface $B$ embedded in a $T= {\rm constant}$ slice, obtain the
data
$\varepsilon$, $j_a$, and $\sigma_{ab}$. In the functional integral for
$\nu[\varepsilon,j,\sigma]$, fix this data on each $t ={\rm constant}$ slice
of $\partial M$. Observe that, to the extent that the physical system can be
approximated by a single classical configuration, that configuration will be
the real stationary black hole that is used to induce the boundary data.

The functional integral (7) can be evaluated semiclassically by searching
for  four--metrics $g_{\mu\nu}$ that extremize $S_m$ and satisfy the specified
boundary conditions. Observe that the Lorentzian black hole geometry that was
used to motivate the choice of boundary conditions is {\it not\/} an extremum
of $S_m$, because it has the topology [Wheeler (spatial)
wormhole]$\times$[time] and cannot be placed on a manifold $M$ with a single
boundary $S^2\times S^1$. However, there is a related complex four--metric
that does extremize $S_m$, and is
described as follows. Let the Lorentzian black hole be given by
$$ds^2 = - \tilde N^2 dT^2 + {\tilde h}_{ij} (dx^i + {\tilde V}^idT) (dx^j +
            {\tilde V}^jdT) \ ,\eqno(8)$$
where $\tilde N$, ${\tilde V}^i$, and ${\tilde h}_{ij}$ are
$T{\hbox{--independent}}$ functions of the spatial coordinates $x^i$.
The horizon coincides with $\tilde N = 0$. Choose spatial
coordinates that are ``co--rotating" with the horizon. Then the proper
spatial velocity of the spatial coordinate system relative to observers at
rest in the $T={\rm constant}$ slices vanishes on the horizon,
$({\tilde V}^i/{\tilde N}) = 0$, and the Killing vector field
$\partial/\partial T$ coincides with the null generator of the horizon.
As shown in [13], the complex metric
$$ds^2 = - (-i\tilde N)^2 dT^2 + {\tilde h}_{ij} (dx^i -i {\tilde V}^idT)
     (dx^j -i    {\tilde V}^jdT) \ ,\eqno(9)$$
where the coordinate $T$ is real, satisfies the Einstein equations
everywhere on a manifold with topology $M=({\hbox{disk}})\times S^2$, with
the possible exception of the points $\tilde N = 0 $ where the foliation
$T={\rm constant}$ degenerates. The locus of those points $\tilde N =0$
is a two--surface called the ``bolt" [18]. Near the bolt, the metric becomes
$$ds^2 \approx {\tilde N}^2 dT^2 + {\tilde h}_{ij} dx^idx^j \ ,\eqno(10)$$
and describes a Euclidean geometry. The sourceless Einstein equations are not
satisfied at the bolt if this geometry has a conical singularity in the
two--dimensional submanifold that contains the unit normals $\tilde n^i$ to
the bolt for each of the $T={\rm constant}$ hypersurfaces. However, there is
no conical singularity if the circumferences of circles surrounding the bolt
initially increase as $2\pi$ times proper radius. The circumference of such
circles is given by $P\tilde N$, where $P$ is the period in coordinate time
$T$. Therefore the absence of conical singularities is insured if the
condition  $P({\tilde n}^i \partial_i\tilde N) = 2\pi$
holds at each point on the bolt, where ${\tilde n}^i$ is the unit normal to
the bolt in one of the $T={\rm constant}$ surfaces. Because the unit normal
is proportional to  $\partial_i\tilde N$ at the bolt, this condition
restricts the period in coordinate time $T$ to be $P=2\pi/\kappa_{\sss H}$,
where $\kappa_{\sss H} = [(\partial_i\tilde N) {\tilde h}^{ij}
(\partial_j\tilde N)]^{1/2}\bigr|_{\sss H}$ is the surface gravity at the
horizon of the Lorentzian black hole (8).

The lapse function and shift vector for the complex metric (9) are
$N=-i\tilde N$ and $V^i = -i{\tilde V}^i$. Thus, (9) and the Lorentzian
metric (8) differ only by a factor of $-i$ in their lapse functions and
shift vectors. In particular, the three--metric ${\tilde h}_{ij}$ and its
conjugate momentum ${\tilde P}^{ij}$ coincide for the stationary metrics
(8) and (9) [13]. Since the boundary data $\varepsilon$, $j_a$, and
$\sigma_{ab}$ are constructed from the canonical variables only, the
complex metric (9) satisfies the boundary conditions imposed on the
functional integral for $\nu[\varepsilon,j,\sigma]$.

The complex metric (9) with the periodic identification given above
extremizes the action $S_m$ and satisfies the chosen boundary conditions
for the density of states $\nu[\varepsilon,j,\sigma]$. Although this metric
is not included in the sum over Lorentzian geometries (7), it can be used
for a steepest descents approximation to the functional integral by
distorting the contours of integration for the lapse $N$ and shift $V^i$
in the complex plane. Then the density of states is approximated by
$$\nu[\varepsilon,j,\sigma] \approx \exp(iS_m[-i\tilde N,-i\tilde V,
     \tilde h])  \ ,\eqno(11)$$
where $S_m[-i\tilde N,-i\tilde V,\tilde h]$ is the microcanonical action
evaluated at the complex extremum (9). The density of states can be
expressed approximately as
$$\nu[\varepsilon,j,\sigma] \approx \exp(\S[\varepsilon,j,\sigma])
     \ ,\eqno(12)$$
where $\S[\varepsilon,j,\sigma]$ is the entropy of the system. Then the
result (11) shows that the entropy is
$$\S[\varepsilon,j,\sigma] \approx iS_m[-i\tilde N,-i\tilde V,\tilde h]
       \eqno(13)$$
for the gravitational field with microcanonical boundary conditions.

In order to evaluate $S_m$ for the metric (9), we start with the
microcanonical action written in spacetime covariant form [8] and
perform a canonical decomposition under the assumption that the
manifold $M$ has the topology of a punctured disk $\times S^2$. That is,
the spacelike hypersurfaces $\Sigma$ have topology $I\times S^2$, and
the timelike direction is periodically identified ($S^1$). The outer
boundary of the disk corresponds to the three--boundary element $\B3$
of $M$ (denoted $\partial M$ previously) on which the boundary values
of $\varepsilon$, $j_a$, and $\sigma_{ab}$ are imposed. The inner
boundary of the disk, the boundary of the puncture, appears as another
boundary element $\Hthree$ for $M$. (No data are specified at
$\Hthree$.) The canonical decomposition results in [8]
$$S_m = \int_M d^4x \bigl[ P^{ij} {\dot h}_{ij} - N\H - V^i\Hi \bigr]
     + \int_\Hthree d^3x\sqrt{\sigma} \bigl[ n^i(\partial_iN)/\kappa +
           2n_iV_jP^{ij}/\sqrt{h}   \bigr]    \ ,\eqno(14)$$
where the expression $a_i = (\partial_iN)/N$ for the acceleration of the
timelike unit normal has been used. The boundary term at $\Hthree$ was
first given in [13].

Now evaluate the action $S_m$ on the punctured disk $\times S^2$ for the
complex metric (9), and take the limit as the puncture disappears to
obtain a manifold topology $M=({\rm disk})\times S^2$. In this limit,
the smoothness of the complex geometry is assured by the regularity
condition on the period of $T$. Since the metric satisfies the Einstein
equations, the Hamiltonian and momentum constraints vanish, and the
terms $P^{ij} {\dot h}_{ij}$ also vanish by stationarity. Moreover, the
second boundary term at $\Hthree$ is zero because the shift vector
vanishes at the horizon. Thus, only the first of the boundary terms at
$\Hthree$ survives. Evaluating this term for the complex metric (9),
that is, for the lapse function $N=-i\tilde N$, and using the regularity
condition $P = 2\pi/\kappa_{\sss H}$, we find for the microcanonical
action
$$S_m[-i\tilde N,-i\tilde V,\tilde h] = -{i\over\kappa} \int_0^P
        dT\int d^2x  \sqrt{\tilde\sigma} {\tilde n}^i\partial_i\tilde N
       = -{i\over4} A_{\sss H}  \ ,\eqno(15)$$
where $A_{\sss H}$ is the area of the event horizon for the Lorentzian
black hole (8).

The result (15) for the microcanonical action evaluated at the extremum
(9) leads to the following approximation for the entropy (13):
$$\S[\varepsilon,j,\sigma] \approx {1\over4}A_{\sss H} \ .\eqno(16)$$
The generality of the result (16) should be emphasized: The boundary
data $\varepsilon$, $j_a$, and $\sigma_{ab}$ were chosen from a general
stationary, axisymmetric black hole that solves the vacuum Einstein
equations within a spatial region with boundary $B$. Outside the
boundary $B$, the black hole spacetime need not be free of matter or
be asymptotically flat. Thus, for example, the black hole can be
distorted relative to the standard Kerr family. Furthermore, recall
that the quantum--statistical system with this boundary data is
classically approximated by the physical black hole solution that
matches that boundary data. The result (16) shows that the entropy of
the system is approximately $1/4$ the area of the event horizon of the
physical black hole configuration that classically approximates the
contents of the system. It also should be emphasized that the entropy
is independent of the term $S^0$ in (1) as has been shown
in the framework of the canonical partition function [9].

Expresson (16) is the principal result we wished to demonstrate here.
The physical and mathematical limitations of our analysis and possible
ways to overcome them are described in [8], which also discusses
canonical and grand canonical boundary conditions for the rotating
black hole. A further elaboration of the physical underpinnings and
implications of our analysis of relativistic rotating systems is
given in [19].

The authors gratefully acknowledge research support received from the
National Science Foundation, grant number PHY--8908741.


{\bf REFERENCES}
\parindent=22pt

\item{[1]}Y. Choquet--Bruhat and J.W. York, The Cauchy Problem,
pp.~99--172 in {\it General Relativity and Gravitation\/}, edited by
A. Held (Plenum Press, New York, 1980).

\item{[2]}W. Israel and J.M. Stewart, in {\it General Relativity and
Gravitation.~II\/}, edited by A. Held (Plenum Press, New York, 1980).

\item{[3]}V. Frolov and K.S. Thorne, Phys. Rev. {\bf D39}, 2125 (1989).

\item{[4]}H.B. Callen, {\it Thermodynamics\/} (Wiley, New York, 1985).

\item{[5]}C. Lanczos, {\it The Variational Principles of Mechanics\/}
(University of Toronto Press, Toronto, 1970).

\item{[6]}J.D. Brown and J.W. York, Phys. Rev. {\bf D40}, 3312 (1989).

\item{[7]}R.P. Feynman and A.R. Hibbs, {\it Quantum Mechanics and Path
   Integrals\/} (Mc\-Graw--Hill, New York, 1965).

\item{[8]}J.D. Brown and J.W. York, Phys. Rev. {\bf D47}, 1420 (1993).

\item{[9]}J.W. York, Phys. Rev. {\bf D33}, 2092 (1986).

\item{[10]}B.F. Whiting and J.W. York, Phys. Rev. Lett. {\bf 61},
1336 (1988).

\item{[11]}H.W. Braden, J.D. Brown, B.F. Whiting, and J.W. York,
Phys. Rev. {\bf D42}, 3376 (1990).

\item{[12]}J.D. Brown and J.W. York, Phys. Rev. {\bf D47}, 1407 (1993).

\item{[13]}J.D. Brown, E.A. Martinez, and J.W. York, Phys. Rev. Lett.
{\bf 66}, 2281 (1991).

\item{[14]}J.D. Brown, G.L. Comer, E.A. Martinez, J. Melmed, B.F.
Whiting, and J.W. York, Class. Quantum Grav. {\bf 7}, 1433 (1990).

\item{[15]}J.D. Brown and J.W. York, ``Jacobi's action and the density
of states", to appear in {\it Festschrift for Dieter Brill\/}, edited
by B.L. Hu and T. Jacobson (Cambridge University Press, Cambridge,
1993).

\item{[16]}J.W. York, Found. Phys. {\bf 16}, 249 (1986).

\item{[17]} G.T. Horowitz, Class. Quantum Grav. {\bf 8}, 587 (1991).

\item{[18]} G.W. Gibbons and S.W. Hawking, Commun. Math. Phys.
{\bf 66}, 291 (1979).

\item{[19]}J.D. Brown and J.W. York, ``Hamiltonian and boundary
conditions for rotating general--relativistic equilibrium systems",
to be published.
\bye